\documentclass[journal]{IEEEtran}
\usepackage{cite}
\usepackage{graphicx,amssymb,lineno}
\usepackage{amsmath,amsfonts,amssymb}

\usepackage{algorithm}
\usepackage{algorithmic}
\usepackage[usenames]{color}
\usepackage{float}
\usepackage{mathtools}
\usepackage{cite}
\usepackage{bm}

\usepackage{graphicx,graphics,color,epsfig,subfigure,graphpap,rotate}
\usepackage{times, verbatim, subfigure, epsfig, graphicx, latexsym, amsmath}
\usepackage{url}
\usepackage{subfigure}
\usepackage{CJK}
\allowdisplaybreaks[4]
\begin{document}
\title{Energy-Constrained Computation Offloading in Space-Air-Ground Integrated Networks using Distributionally Robust Optimization}

\author{Yali~Chen,
        Bo~Ai,~\IEEEmembership{Senior Member,~IEEE},
        Yong~Niu,~\IEEEmembership{Member,~IEEE},
        Hongliang~Zhang,~\IEEEmembership{Member,~IEEE},
        and Zhu~Han,~\IEEEmembership{Fellow,~IEEE}

\thanks{Copyright (c) 2015 IEEE. Personal use of this material is permitted. However, permission to use this material for any other purposes must be obtained from the IEEE by sending a request to pubs-permissions@ieee.org. This study was supported in part by the National Key Research and Development Program of China under Grant 2020YFB1806903; in part by the Fundamental Research Funds for the Central Universities under Grant 2020YJS218; in part by the National Natural Science Foundation of China under Grants 61725101, 61961130391, 61801016, and U1834210; in part by the Royal Society Newton Advanced Fellowship under Grant NA191006; in part by the State Key Laboratory of Rail Traffic Control and Safety under Grants RCS2020ZT010, RCS2019ZZ007, and RCS2021ZT009; in part by the Fundamental Research Funds for the Central Universities, China, under Grant 2020JBZD005 and Grant 2020JBM089; in part by the open research fund of National Mobile Communications Research Laboratory, Southeast University (No. 2021D09); in part by Frontiers Science Center for Smart High-speed Railway System; in part by the Project of China Shenhua under Grant (GJNY-20-01-1); and in part by US NSF CNS-2128368, CNS-2107216, Toyota and Amazon. (\emph{Corresponding authors: B. Ai, Y. Niu.})}

\thanks{Y. Chen is with the State Key Laboratory of Rail Traffic Control and Safety, Beijing Jiaotong University, Beijing 100044, China, and also with the Frontiers Science Center for Smart High-Speed Railway System, Beijing Jiaotong University, Beijing 100044, China (e-mail: chenyali@bjtu.edu.cn).}

\thanks{B. Ai is with the State Key Laboratory of Rail Traffic Control and Safety, Beijing Jiaotong University, Beijing 100044, China, and also with Henan Joint International Research Laboratory of Intelligent Networking and Data Analysis, Zhengzhou University, Zhengzhou 450001, China, and also with Research Center of Networks and Communications, Peng Cheng Laboratory, Shenzhen, China, and also with the Frontiers Science Center for Smart High-Speed Railway System, Beijing Jiaotong University, Beijing 100044, China (e-mail: boai@bjtu.edu.cn).}

\thanks{Y. Niu is with the State Key Laboratory of Rail Traffic Control and Safety, Beijing Jiaotong University, Beijing 100044, China, and also with the National Mobile Communications Research Laboratory, Southeast University, Nanjing 211189, China (e-mail: niuy11@163.com).}

\thanks{H. Zhang is with the Department of Electrical Engineering, Princeton University, Princeton, NJ 08544, USA (e-mail: hongliang.zhang92@gmail.com).}

\thanks{Z. Han is with the Department of Electrical and Computer Engineering in the University of Houston, Houston, TX 77004 USA, and also with the Department of Computer Science and Engineering, Kyung Hee University, Seoul, South Korea, 446-701. (e-mail: zhan2@uh.edu).}

}

\maketitle

\begin{abstract}
With the rapid development of connecting massive devices to the Internet, especially for remote areas without cellular network infrastructures, space-air-ground integrated networks (SAGINs) emerge and offload computation-intensive tasks. In this paper, we consider a SAGIN, where multiple low-earth-orbit (LEO) satellites providing connections to the cloud server, an unmanned aerial vehicle (UAV), and nearby base stations (BSs) providing edge computing services are included. The UAV flies along a fixed trajectory to collect tasks generated by Internet of Things (IoT) devices, and forwards these tasks to a BS or the cloud server for further processing. To facilitate efficient processing, the UAV needs to decide where to offload as well as the proportion of offloaded tasks. However, in practice, due to the variability of environment and actual demand, the amount of arrival tasks is uncertain. If the deterministic optimization is utilized to develop offloading strategy, unnecessary system overhead or higher task drop rate may occur, which severely damages the system robustness. To address this issue, we characterize the uncertainty with a data-driven approach, and formulate a distributionally robust optimization problem to minimize the expected energy-constrained system latency under the worst-case probability distribution. Furthermore, the distributionally robust latency optimization algorithm is proposed to reach the sub-optimal solution. Finally, we perform simulations on the real-world data set, and compare with other benchmark schemes to verify the efficiency and robustness of our proposed algorithm.
\end{abstract}


\section{Introduction}\label{S1}
Internet of Things (IoT) devices are becoming more attractive and intensively deployed in various scenarios, e.g., wearable devices, smart phones and home appliances in urban areas, and sensors for environmental monitoring management in remote areas. On the one hand, they are advancing the development of smart environment, such as face recognition, automatic navigation, forest disaster monitoring and management, forest weather change monitoring, forest pest monitoring and forest intelligent fire prevention early warning. On the other hand, these IoT devices have low power and low computing capability, but need to execute computation-intensive tasks, which brings significant challenges \cite{Li1}. To solve these problems, offloading tasks to local edge servers to take full advantage of multi-access edge computing (MEC) and offloading to the cloud server via macro base stations (MBSs) are the potential solutions.

However, cellular networks cannot provide ubiquitous coverage \cite{boai}. In some remote areas with intractable access points deployment, or in some areas with destroyed infrastructures caused by natural disasters, computation offloading to cellular networks becomes impractical, and we need to seek for more effective solutions \cite{Ziye}. Fortunately, space-air-ground integrated networks (SAGINs) have received an ever-increasing level of attention for the future communication systems to provide great strength in the seamless coverage and accommodate diverse services \cite{Kato,Shi}. Specifically, it has a heterogeneous architecture, which includes three network segments, i.e., satellite, aerial and ground network segments.

For the satellite segment, low-earth-orbit (LEO) satellites have been promising options to provide always-on and high-capacity cloud computing services, and further extend the coverage. In practice, SpaceX and OneWeb have been developing LEO satellite constellations at lower altitudes \cite{SpaceX,OneWeb}. Besides, although the cloud server has very attractive computing and processing capabilities, the transmission distance between the LEO satellite and the ground terminal still induces a long propagation latency. For the aerial segment as a temporary support of ground networks, such as unmanned aerial vehicles (UAVs), high-altitude platforms (HAPs), and balloons, they can be despatched as relays or edge processors to assist terrestrial communications on demand \cite{AHMED,Zhang1,Zhang3}. The air-ground channel quality is more likely to be better due to the line-of-sight (LoS) communication, which greatly saves the transmission power and energy consumption of IoT devices, while prolonging their operational lifetime. In addition, the deployment of these communication systems in the aerial segment is flexible and cost-effective \cite{Zhang2}. For the ground segment, it has abundant facilities and can provide access to most requests in urban areas, as well as provide edge computing services indirectly for areas that are not covered. Benefiting from the complementary advantages of three component networks of SAGIN, the reliable and high-speed services are achieved, and the system performance is greatly improved.

In this paper, we consider a large number of IoT devices deployed in remote areas without the cellular network coverage. The UAV in the SAGIN architecture is dispatched to fly along a given trajectory to collect tasks from IoT devices, and then decides to offload tasks to a nearby ground BS server over the C-band or a certain LEO satellite via the Ka-band, and makes a decision about the ratio of offloading tasks and local processing tasks.

However, the amount of tasks gathered by the UAV is highly dynamic \cite{Xu}. In most cases, not only the parameter of task volume is uncertain, but also the probability distribution is unknown and hard to accurately predict. Inherent uncertainty and inaccurate estimation of the probability distribution seriously affect the system reliability. For instance, in the worst case, the estimated value is much smaller than the actual value. If we still use the traditional deterministic optimization, many tasks cannot be processed, which leads to constant retransmissions and a soaring system latency. Thus, a relatively conservative strategy is urgent to ensure the robustness of the system. Recently, distributionally robust optimization (DRO) has received great attention in both the operations research and statistical learning communities \cite{Delage}. As a data-driven method, it can characterize the uncertainty of probability distribution of arrival tasks by constructing the confidence set, which is actually an alias of the uncertainty set and consists a family of ambiguous probability distributions \cite{Zhao3}. In other words, the underlying true distribution is ambiguous to decision makers. According to the observed historical data, the confidence set can be adjusted with the confidence level guaranteed. Compared with the traditional deterministic optimization, stochastic programming with specific probability distribution, and robust optimization \cite{Gorissen} with the probabilistic information ignored, DRO better balances the system performance and the conservativeness.

For these aggregated tasks, the latency is of paramount importance. Combining the unknown probability distribution of arriving tasks, we present several metrics to construct the confidence set and formulate the distributionally robust optimization problem of minimizing the expected system latency under the worst-case probability distribution. Then, the optimization problem is solved with the proposed distributionally robust
latency optimization algorithm and the sub-optimal solution is yielded. The contributions of this paper are summarized as follows.

\begin{itemize}
\item Without the presumption of probability distribution of arrival tasks from IoT devices, we utilize the reference distribution generated by historical observation data and different distance metrics to construct a confidence set. Then, we formulate the problem as a distributionally robust optimization problem, where the worst-case expected system latency is minimized over a set of probability distributions with the total energy consumption constrainted.

\item The optimization problem is a mixed integer programming problem. Optimization variables include the UAV's access decision and the allocation of offloading tasks at each time slot. The former one indicates whether the UAV accesses to a nearby BS or a satellite to further connect to the cloud server, and the latter one is the proportion of offloaded tasks. Then, we propose the distributionally robust latency optimization algorithm to obtain a sub-optimal and robust solution within the confidence set.

\item Based on the real-world data set, we evaluate the system performance comparing with traditional deterministic optimization and other benchmark schemes to verify the efficiency and the robustness. Meanwhile, different metrics for the confidence set are also discussed.
\end{itemize}

The rest of the paper is organized as follows. In Section~\ref{S2}, we summarize the related work of deterministic optimization and optimization under uncertainty. In Section~\ref{S3}, the system model with a SAGIN architecture is presented. In Section~\ref{S4}, we utilize a statistical inference method to construct the confidence set, and formulate an energy-constrained latency minimization problem based on the distributionally robust optimization framework. In Section~\ref{S5}, we propose a distributionally robust latency optimization algorithm to solve this problem and obtain the sub-optimal solution. Performance evaluation on the real-world data set is given in Section~\ref{S6}. Finally, we conclude this paper in Section~\ref{S7}.

\section{Related Work}\label{S2}
There have been several related works about deterministic optimization in air-ground, satellite-ground and SAGIN communications. Yang \emph{et al.} \cite{Yang} studied a system that UAV was enabled to collect an amount of data from its served ground terminal (GT), and then optimized the UAV trajectory and transmitted power of the GT to reach a tradeoff between the uplink communication energy of GT and the propulsion energy consumption of UAV. Ren \emph{et al.} \cite{Ren} considered a UAV-enabled relay system to assist GTs to deliver messages with ultra-high reliability and low latency, and then studied UAV location optimization and power allocation. Hammouti \emph{et al.} \cite{Hammouti} designed the joint users-UAVs matching, 2D placement, and dynamic altitude adjustment of UAVs, for the purpose of maximizing the system sum rate with the quality of service guaranteed. Zhou \emph{et al.} \cite{ZhouYi} presented a UAV-enabled communication, computing and caching virtual reality (VR) delivery system, and jointly optimized UAV location, backhaul and fronthaul bandwidth, computing capacity, caching and computing policies. Sun \emph{et al.} \cite{SunZhou} proposed a two-tier network architecture, where access points (APs) collected packets from served IoT devices and delivered aggregated data to the UAV, and then studied the UAV's trajectory design and resource allocation. Di \emph{et al.} \cite{Di} proposed a terrestrial-satellite network with ultra-dense LEO integrated, where the ground user could access the network through the small cell, the macro cell, or the LEO-based small cell with large backhual capacity. By jointly optimizing the ground data offloading, association between terrestrial-satellite terminals and multiple LEO satellites, and resource allocation, the objective of maximizing the number of accessed users and the sum rate of all cells with the backhaul capacity constrainted was achieved. Liu \emph{et al.} \cite{Liu} provided a comprehensive survey about recent research works on SAGINs, where several existing network architectures were presented, and some technical challenges were pointed out. Under a SAGIN scenario, Zhou \emph{et al.} \cite{Zhou} adopted a UAV to make online channel access and offloading decisions for the collected terrestrial IoT tasks to minimize the expectation of long-term average system delay with the system energy budget and UAV's buffer capacity taken into consideration. Cheng \emph{et al.} \cite{Cheng} presented a SAGIN edge/cloud computing framework for remote IoT users. Then, they formulated a joint task scheduling and virtual machines allocation mechanism for UAV edge servers, and investigated a computing offloading problem to minimize the weighted sum of server usage cost, energy consumption and delay in the SAGIN.

In recent years, a considerable research effort has been made in optimization under uncertainty. Bertsimas \emph{et al.} \cite{Bertsimas1} presented a mathematical framework that was suitable for problems with rarely available precise knowledge and captured dynamic decision making under uncertainty. Li \emph{et al.} \cite{Li2} integrated transcoding, proactive caching and backhaul retrieving in a MEC-enabled adaptive streaming system, and formulated a data-driven risk-averse optimization for delivery scheduling and caching strategy with an uncertain distribution of video request arrivals. Wang \emph{et al.} \cite{Wang} employed the data-driven approach to study the trade-off between the temporal operational privacy of primary users by adding dummy signals and the service requirements of secondary users with uncertain traffic demands. Zhao \emph{et al.} \cite{Zhao1} introduced a data-driven risk-averse stochastic optimization framework and studied the Wasserstein metric for both continuous and discrete cases of uncertainty distribution. Rahimian \emph{et al.} \cite{Rahimian} provided an interpretation of DRO and its connections with other concepts, such as function regularization, robust optimization, chance-constrained and risk-aversion optimization, and game theory. Besides, two types of techniques for solving DRO problems were reviewed, and different models used to represent the ambiguity set of distributions were discussed. Chen \emph{et al.} \cite{Chen} exploited differential privacy preserving protocols for users' historical demand database, and employed the risk-averse two-stage stochastic programming optimization to model the small cell planning problem with the demand uncertainty characterized. The optimization objective was to minimize the capital and operating costs for micro operators. Zhao \emph{et al.} \cite{Zhao2} introduced the $\zeta$-structure probability metric family and explored the relationship between different types of metrics utilized to construct the confident set. According to the confident set, they developed the solution methodologies to solve the risk-averse two-stage stochastic program for the continuous and discrete true distribution cases.

\begin{figure}[!t]
\begin{center}
\includegraphics*[width=0.7\columnwidth,height=2.3in]{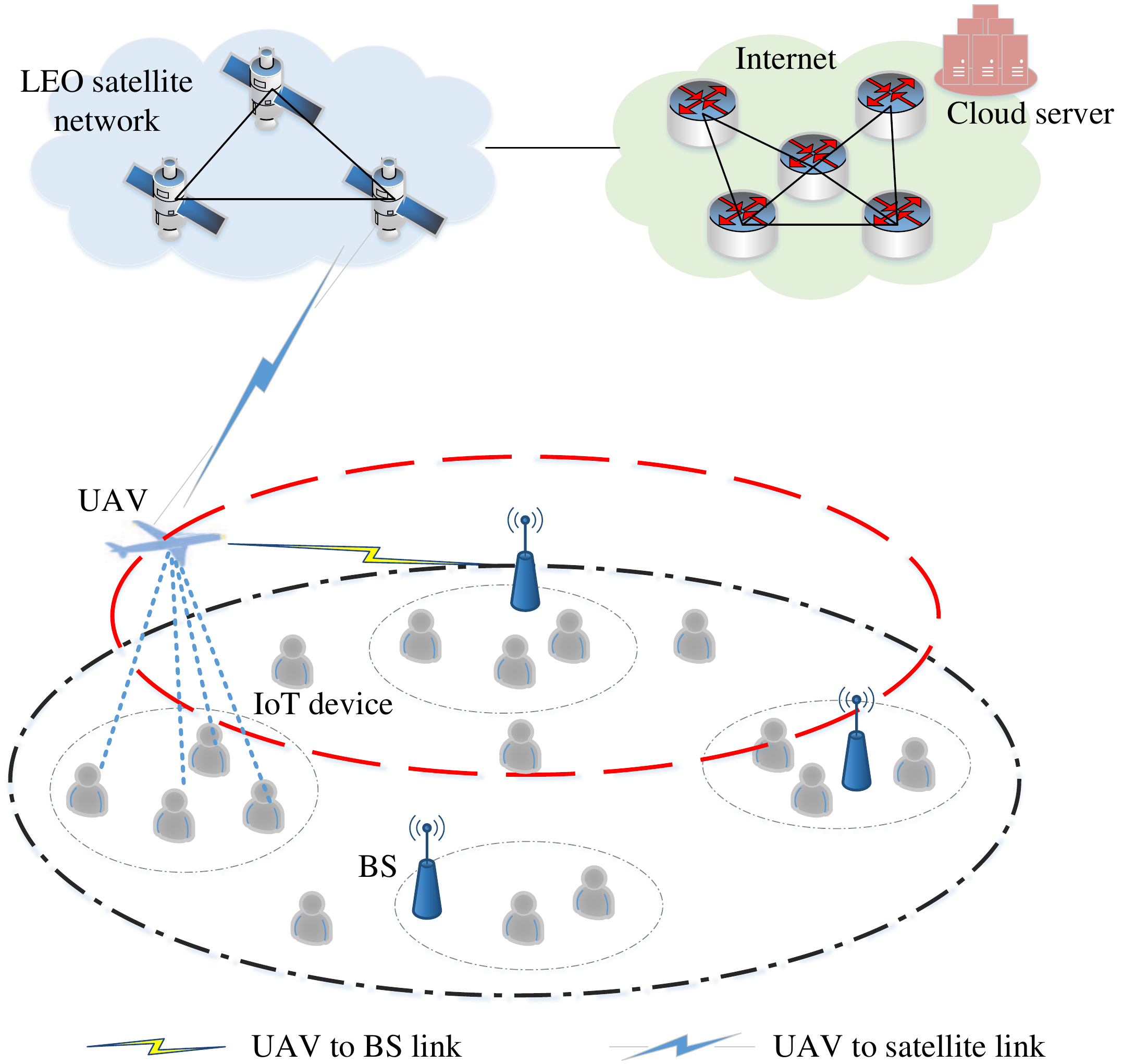}
\end{center}
\caption{System model for a SAGIN architecture.} \label{fig1}
\end{figure}

\section{System Model}\label{S3}
For IoT devices deployed in the areas without cellular network coverage, we introduce a SAGIN, as shown in Fig. \ref{fig1}, to handle service requests from dense IoT devices. In the studied time horizon, the timeline is divided into slots. $T$ is the total number of time slots and $\tau$ is the duration of each time slot. At the beginning of each time slot $t$, a fixed-wing UAV flies with a constant speed of $v$ to collect data tasks in these areas. The radius of the scheduled circular horizontal trajectory is $r$ and the height is $H$. Then, we assume that there are a series of optional surrounding BSs $\textbf{B}=\{1,2,...,N\}$, and $\textbf{S}=\{1,2,...,M\}$ is the set of indices of $M$ connectable satellites above these investigated areas.

According to the amount of tasks received, the UAV needs to make a decision about whether to offload to a certain satellite within the visible range, and then utilize the cloud server for processing, or offload to a nearby BS for processing. When the offloaded destination is selected, the proportion of offloaded tasks also needs to be determined. This is because the computing capability of the UAV is still limited. Therefore, a part of tasks are executed by the UAV, and the remaining tasks are processed by the offloaded destination. Then, we introduce the channel models from the UAV to the BS and from the UAV to the satellite.

The channel between UAV $u$ and BS $b$ works in the C-band, and channel coefficient $h_{ubt}$ at time slot $t$ can be expressed as
\begin{equation}
h_{ubt}=h_{0}\beta_{0}\sqrt{PL(d_{ubt})},
\label{eq1}
\end{equation}
where $h_{0}$ is the small-scale fading following the Rayleigh distribution, i.e., $h_{0}\sim \mathcal{CN}(0,1)$, $\beta_{0}$ is the shadow fading, $d_{ubt}$ is the distance between the UAV and BS $b$ in time slot $t$, and $PL(d_{ubt})$ is the large-scale pathloss. Although the communication distance between the UAV and BS is time-varying, we assume that the position of the UAV is static in one time slot, which is the one at the end of each time slot for brevity.

Next, the transmitted rates of links in bits/second from the UAV to the BS and from the BS to the UAV at time slot $t$ are given as follows, repsectively,
\begin{equation}
R_{ubt}=B_{ub}\log_{2}\left(1+\frac{P_{ub}|h_{ubt}|^{2}}{N_{0}B_{ub}}\right),
\label{eq2}
\end{equation}
\begin{equation}
R_{but}=B_{ub}\log_{2}\left(1+\frac{P_{bu}|h_{ubt}|^{2}}{N_{0}B_{ub}}\right),
\label{eq3}
\end{equation}
where $B_{ub}$ is the allocated bandwidth, $P_{ub}$ is the transmitted power from the UAV to the BS, and $P_{bu}$ is the transmitted power in the opposite direction. In addition, $N_{0}$ is the Gaussian noise power spectral density of the terrestrial network.

For the transmission link from the UAV to the satellite, the LoS path is clear and dominant. Together with weak scatter components propagated via different non-LoS (NLoS) paths, we model the UAV-satellite channel as a Rician channel \cite{Di,Abdi}. The channel gain, which includes the LoS and NLoS components, is shown as
\begin{equation}
h_{ust}=\sqrt{\frac{\eta}{1+\eta}}h^{L}_{ust}+\sqrt{\frac{1}{1+\eta}(d_{ust})^{-\alpha^{\prime}}}h^{NL}_{ust},
\label{eq4}
\end{equation}
where $\eta$ is the Rician factor, and $h^{L}_{ust}$ is the channel gain in the LoS case, which can be represented by
\begin{equation}
h^{L}_{ust}=\sqrt{(d_{ust})^{-\alpha}}e^{-j\frac{2\pi}{\lambda}d_{ust}}.
\label{eq5}
\end{equation}
Here, $d_{ust}$ is the distance between the UAV and the selected satellite $s$ at $t$, $\alpha$ is the pathloss exponent in the LoS case, and $\lambda$ is the wavelength of Ka-band. Since the orbit of each satellite is pre-designed, its longitude, latitude, and altitude information are known in each time slot, and we assume that the position remains unchanged during the time period. $\alpha^{\prime}$ is the pathloss exponent under the NLoS transmission, and $h^{NL}_{ust}$ is the corresponding small-scale channel gain, which follows that $h^{NL}_{ust}\sim \mathcal{CN}(0,1)$.

Then, the achievable transmission rates from the UAV to the satellite denoted as $R_{ust}$ and from the satellite to the UAV denoted as $R_{sut}$ can be accordingly derived as
\begin{equation}
R_{ust}=B_{us}\log_{2}\left(1+\frac{P_{us}G_{0}|h_{ust}|^{2}}{N_{0}B_{us}}\right),
\label{eq6}
\end{equation}
\begin{equation}
R_{sut}=B_{us}\log_{2}\left(1+\frac{P_{su}G_{0}|h_{ust}|^{2}}{N_{0}B_{us}}\right),
\label{eq7}
\end{equation}
where $B_{us}$, $P_{us}$, $P_{su}$ and $G_{0}$ represent the bandwidth of Ka-band, the transmission power from the UAV to the satellite, the transmission power from the satellite to the UAV, and fixed antenna gain.

\section{Problem Formulation}\label{S4}
In this section, with the uncertainty in the amount of tasks collected by the UAV considered, we describe how to use the data-driven approach to construct the confidence set, and provide the distributionally robust latency optimization problem formulation with the system energy consumption constrained.

\subsection{Confidence Set Construction}\label{S4-1}
In this paper, the task request received by the UAV is uncertain, and we have not given a definite probability distribution followed by the arriving tasks. Instead, a confidence set is built for ambiguous distributions using distance metrics to directly associate the robustness of the model with the observed historical data. Furthermore, the confidence set is data-driven and it is assumed to contain the true distribution.

Supposing that there is a compact convex sample space $\Omega$, which contains $K$ possible discrete values of the task volume $\Omega=\{\xi_{1},\xi_{2},...,\xi_{K}\}$, and the dimension of the sample space is 1. For a series of obtained historical data with a total of $K^{\prime}$, the reference distribution is expressed as $P_{0}=\{p^{0}_{1}, p^{0}_{2},...,p^{0}_{K}\}$. In this paper, we use the empirical distribution as the reference distribution. Since the supporting space is discrete, the empirical distribution is just like a step function that jumps $1/K^{\prime}$ at each sample point, that is, $p^{0}_{k}=\frac{1}{K^{\prime}}\sum\limits_{i=1}^{K^{\prime}}\delta_{\xi^{\prime}_{i}}(\xi_{k}), \forall k=1,2,...,K$. $\delta_{\xi^{\prime}_{i}}(\xi_{k})=1$ only when $\xi^{\prime}_{i}=\xi_{k}$ and otherwise, $\delta_{\xi^{\prime}_{i}}(\xi_{k})=0$. Then, we let the true probability distribution ambiguous and denote it as $P=\{p_{1}, p_{2},...,p_{K}\}$ corresponding to parameters in $\Omega$.

Next, considering that the reference distribution is not necessarily equivalent to the actual distribution, we measure the distance between the ambiguity distribution $P$ and the reference distribution $P_{0}$, and construct a confidence set for the ambiguous distribution, represented as $D$. However, different metrics usually make the measured distance between two distributions different, and then have a significant impact on the effectiveness and robustness of the system. Thus, we choose three metrics for discussion in this paper. First of all, $L_{1}$ norm and $L_{\infty}$ norm metrics are adopted because of their excellent numerical tractability. The distance measurements for $L_{1}$ norm and $L_{\infty}$ norm, denoted as $d_{1}(P,P_{0})$ and $d_{\infty}(P,P_{0})$, are defined as follows,
\begin{equation}
d_{1}(P,P_{0})=\parallel P-P_{0}\parallel_{1}=\sum\limits_{k=1}^{K}|p_{k}-p^{0}_{k}|,
\label{eq8}
\end{equation}
\begin{equation}
d_{\infty}(P,P_{0})=\parallel P-P_{0}\parallel_{\infty}=\max\limits_{1\leq k \leq K}|p_{k}-p^{0}_{k}|.
\label{eq9}
\end{equation}

Then, the statistical inference method is utilized to construct the confidence set. Corresponding to $L_{1}$ norm and $L_{\infty}$ norm, we additionally introduce two parameters, $\theta_{1}$ and $\theta_{\infty}$, to construct confidence sets $D_{1}$ and $D_{\infty}$, respectively given by
\begin{equation}
\begin{split}
D_{1}&=\left\{P\in R^{K}_{+}|\parallel P-P_{0}\parallel_{1}\leq \theta_{1} \right\}\\
&=\left\{P\in R^{K}_{+}|\sum\limits_{k=1}^{K}|p_{k}-p^{0}_{k}|\leq \theta_{1} \right\},
\label{eq10}
\end{split}
\end{equation}
\begin{equation}
\begin{split}
D_{\infty}&=\{P\in R^{K}_{+}|\parallel P-P_{0}\parallel_{\infty}\leq \theta_{\infty} \}\\
&=\left\{P\in R^{K}_{+}|\max\limits_{1\leq k \leq K}|p_{k}-p^{0}_{k}|\leq \theta_{\infty} \right\},
\label{eq11}
\end{split}
\end{equation}
where $\theta_{1}$ and $\theta_{\infty}$ are tolerance values and closely related to the confidence level $\beta$ and the size of available historical data $K^{\prime}$, and we try to get the precise relationship between them. As indicated in \cite{Zhao4}, the convergence rates between the ambiguous distribution $P$ and the reference distribution $P_{0}$ under $L_{1}$ norm and $L_{\infty}$ norm are shown as follows, respectively,
\begin{equation}
Pr\{\parallel P-P_{0}\parallel_{1}\leq \theta_{1}\}\geq 1-2K\exp(-2K^{\prime}\theta_{1}/K),
\label{eq12}
\end{equation}
\begin{equation}
Pr\{\parallel P-P_{0}\parallel_{\infty}\leq \theta_{\infty}\}\geq 1-2K\exp(-2K^{\prime}\theta_{\infty}).
\label{eq13}
\end{equation}
Let $1-2K\exp(-2K^{\prime}\theta_{1}/K)=\beta$ and $1-2K\exp(-2K^{\prime}\theta_{\infty})=\beta$, as a consequence, $\theta_{1}=\frac{K}{2K^{\prime}}\ln\frac{2K}{1-\beta}$ for the $L_{1}$ norm metric and $\theta_{\infty}=\frac{1}{2K^{\prime}}\ln\frac{2K}{1-\beta}$ for the $L_{\infty}$ norm metric. In other words, the probability of the ambiguous distribution within this confidence set is at least equal to a certain confidence level $\beta$, which is guaranteed. Besides, we can see that $\theta_{1}$ and $\theta_{\infty}$ become smaller as the size of historical data increases, which leads to a tighter confidence set and a less conservative output solution. Thus, $\theta_{1}$ and $\theta_{\infty}$ provide an adjustable bound for the distance between the reference distribution and the ambiguous distribution. Last but not least, as the historical data size increases to infinity, both $\theta_{1}$ and $\theta_{\infty}$ decrease to zero. Then, the confident set becomes a singleton, and the reference distribution can be used as the final true distribution.

In addition, the $\zeta$-structure probability metrics are used to quantify the distance between the reference distribution $P_{0}$ and the ambiguous distribution $P$, which are defined by
\begin{equation}
d_{\zeta}(P,P_{0})=\sup\limits_{h\in \mathcal{H}}\left|\displaystyle{\int_{\Omega}hdP-\int_{\Omega}hdP_{0}}\right|,
\label{eq14}
\end{equation}
where $\mathcal{H}$ is a family of real-valued bounded measurable functions on $\Omega$, and different family members of probability metrics are distinguished according to the definition of $\mathcal{H}$. The Kantorovich metric, as a member of the $\zeta$-structure metrics family, has many applications in the bioinformatics, data mining, transportation theory, and etc. Thus, we explore it in this paper. We denote the distance measure of the Kantorovich metric as $d_{K}(P,P_{0})$. $\mathcal{H}=\{h:\parallel h\parallel_{L}\leq 1\}$, where $\parallel h\parallel_{L}:=\sup\{(h(x)-h(y))/\rho(x,y):x\neq y$ in $\Omega\}$. It should be noted that $\rho(x,y)$ is the distance between two random variables $x$ and $y$.

Then, given a tolerance value $\theta_{kan}$, we can get the form of the confidence set under the Kantorovich metric as follows,
\begin{equation}
\begin{split}
D_{kan}&=\{P\in R^{K}_{+}|\parallel P-P_{0}\parallel_{kan}\leq \theta_{kan} \}\\
&=\left\{P\in R^{K}_{+}|\max\limits_{h\in \mathcal{H}}\left|\displaystyle{\int_{\Omega}hdP-\int_{\Omega}hdP_{0}}\right|\leq \theta_{kan} \right\}.
\label{eq15}
\end{split}
\end{equation}
One particularly important point is that any metric used to construct a confidence set should guarantee that the reference distribution can converge to the true distribution with the size of historical data reaches infinity. As for the convergence of the Kantorovich metric, by adopting the Dvoretzky-Kiefer-Wolfowitz inequality according to \cite{Zhao2}, we can derive as follows,
\begin{equation}
Pr\{\parallel P-P_{0}\parallel_{kan}\leq \theta_{kan}\}\geq 1-\exp(-\theta^{2}_{kan}K^{\prime}/2K^{2}),
\label{eq16}
\end{equation}
where the right side of the inequality is actually our confidence level $\beta$, representing the lower bound of the probability that the ambiguity distribution exists in this given confidence set. Thus, we can get $1-\exp(-\theta^{2}_{kan}K^{\prime}/2K^{2})=\beta$ and $\theta_{kan}=K\sqrt{\frac{2}{K^{\prime}}\ln\frac{1}{1-\beta}}$. There is no doubt that the reference distribution will eventually converge to the true distribution with more historical data observed. When the confidence level remains the same, as the size of historical data increases, we can prove that $\theta_{kan}$ becomes smaller and the confidence set $D_{kan}$ shrinks, which results in a closer distance between the reference distribution and the true distribution.

\subsection{Distributionally Robust Latency Optimization Problem Formulation}\label{S4-2}
In each time slot $t$, we assume that the arriving workload is $\xi_{k}\in \Omega$. Then, we define two binary variables $x_{bt}$ and $x_{st}$ to respectively indicate whether BS $b$ or LEO satellite $s$ is selected as the processor of partial offloaded tasks. In practice, the UAV can only offload tasks to one satellite or BS, i.e., $\sum\limits_{b=1}^{N}x_{bt}+\sum\limits_{s=1}^{M}x_{st}=1$. Since the UAV, BS and satellite are all capable of performing tasks, in what follows, when a part of tasks are processed locally in the UAV, or there is a $b\in \textbf{B}$ that makes $x_{bt}=1$ or there is a $s\in \textbf{S}$ that makes $x_{st}=1$, we analyze the corresponding energy consumption and latency.

Based on the total amount of arriving tasks $\xi_{k}$ in the current time slot $t$, we denote the amount of tasks that the UAV needs to process as $y_{ut}(\xi_{k})$. Therefore, the latency is mainly caused by the computing of the UAV, which is given by
\begin{equation}
L_{ut}=\frac{\delta y_{ut}(\xi_{k})}{f_{u}},
\label{eq17}
\end{equation}
where $f_{u}$ is the UAV computing capability with the unit of cycles per second and $\delta$ is the computation workload to data ratio in cycles/bit depending on the type of the application. Furthermore, the energy expenditure on supporting the mobility of the UAV and maintaining it aloft is not trivial. Here, we take this component into consideration and give the total propulsion energy consumption of $T$ time slots as
\begin{equation}
E^{fly}_{u}=\left[\left(c_{1}+\frac{c_{2}}{g^{2}r^{2}}\right)v^{3}+\frac{c_{2}}{v}\right]T\tau,
\label{eq18}
\end{equation}
where $c_{1}$ and $c_{2}$ are two parameters related to the air density, wing area, aircraft's weight, and so on. $g$ is the gravitational acceleration with nominal value $9.8 m/s^{2}$ \cite{Zeng}.

For the case where a connection is established between the UAV and the BS for task offloading, the total latency overhead consists of the transmission latency from the UAV to the BS, the BS processing latency, and the latency of returning computing results, which is expressed as
\begin{equation}
L_{ubt}=\frac{y_{bt}(\xi_{k})}{R_{ubt}}+\frac{\delta y_{bt}(\xi_{k})}{f_{b}}+\frac{\varpi y_{bt}(\xi_{k})}{R_{but}},
\label{eq19}
\end{equation}
where $y_{bt}(\xi_{k})$ is the amount of assigned tasks and $f_{b}$ is the CPU processing speed of the BS server. After the task computation, the amount of the returned data is generally small but still contributes a non-negligible latency. Thus, we use $\varpi$ to represent the proportional coefficient of returned data size. Besides, the energy consumption in this case only involves the forward transmission energy consumption, while the energy consumption in the computing process and returned transmission are omitted based on practical considerations. The following is the obtained energy consumption:
\begin{equation}
E_{ubt}=\frac{P_{ub}y_{bt}(\xi_{k})}{R_{ubt}}.
\label{eq20}
\end{equation}

For the case where a connection is established between the UAV and a LEO satellite for task offloading, the calculation of latency and energy consumption will be more complicated. The total latency, written as $L_{ust}$, includes the round-trip transmission latency from the UAV to the satellite, the round-trip latency from the satellite to the cloud server, as well as the computing-related latency for the cloud server to process data, i.e.,
\begin{equation}
L_{ust}=\frac{y_{st}(\xi_{k})}{R_{ust}}+\frac{y_{st}(\xi_{k})+\varpi y_{st}(\xi_{k})}{R_{sc}}+\frac{\delta y_{st}(\xi_{k})}{f_{c}}+\frac{\varpi y_{st}(\xi_{k})}{R_{sut}},
\label{eq21}
\end{equation}
where $y_{st}(\xi_{k})$ is the amount of assigned tasks, $R_{sc}$ is the transmission rate from the satellite to the cloud that we simplify to a constant, and $f_{c}$ represents the processing capacity of the cloud server. The energy consumption, denoted as $E_{ust}$, generated in this case is only the transmission energy consumption from the UAV to the cloud server, which is given as follows,
\begin{equation}
E_{ust}=\frac{P_{us}y_{st}(\xi_{k})}{R_{ust}}+\frac{P_{s}y_{st}(\xi_{k})}{R_{sc}}.
\label{eq22}
\end{equation}

Based on the distributionally robust optimization model, we formulate our optimization problem, whose objective is to minimize the expected total latency of $T$ time slots under the worst-case distribution realization in $D$. Here, $D$ can be constructed according to the used metric. Then, the formulated problem is written as
\begin{align}\label{eq23}
&\min\limits_{\textbf{X},\textbf{Y}}\max\limits_{P\in D}\mathbb{E}_{P}[\psi(\textbf{X},\textbf{Y},\xi_{k})] \nonumber \\
& s.t. \nonumber \\
& (a)\ x_{bt} \in \{0,1\}, \ x_{st} \in \{0,1\}, \  \forall  b,s,t, \nonumber \\
& (b)\ \sum\limits_{b=1}^{N}x_{bt}+\sum\limits_{s=1}^{M}x_{st}=1, \  \forall  t, \nonumber \\
& (c)\ y_{ut}(\xi_{k})+\sum\limits_{b=1}^{N}y_{bt}(\xi_{k})+\sum\limits_{s=1}^{M}y_{st}(\xi_{k})=\xi_{k}, \ \forall  t,k, \nonumber \\
& (d)\ y_{ut}(\xi_{k}), y_{bt}(\xi_{k}), y_{st}(\xi_{k})\geq 0, \ \forall  b,s,t,k, \nonumber\\
& (e)\ y_{ut}(\xi_{k})\!\leq \!C_{u}, y_{bt}(\xi_{k})\!\leq \!C_{b}x_{bt}, y_{st}(\xi_{k})\!\leq \!C_{s}x_{st}, \ \forall  b,s,t,k, \nonumber\\
& (f)\  \psi(\textbf{X},\textbf{Y},\xi_{k})\!=\!\sum\limits_{t=1}^{T}\max\left(\frac{\delta y_{ut}(\xi_{k})}{f_{u}},\sum\limits_{b=1}^{N}L_{ubt}\!+\!\sum\limits_{s=1}^{M}L_{ust}\right), \nonumber \\
& (g)\ E^{fly}_{u}+\sum\limits_{b=1}^{N}\sum\limits_{t=1}^{T}E_{ubt} +\sum\limits_{s=1}^{M}\sum\limits_{t=1}^{T}E_{ust}\leq E_{max}, \ \forall  k.
\end{align}
Constraint $(a)$ represents two binary variables $x_{bt}$ and $x_{st}$. Constraint $(b)$ states that the UAV can only offload tasks to one server at a time slot. Constraints $(c)$ and $(d)$ indicate that at each time slot, the sum of tasks processed by the UAV, all BSs, and all LEO satellites equals to the amount of tasks arrived, and the tasks allocated to any individual processor are non-negative. A very important point is to explain the linear inequality in constraint $(e)$. $y_{ut}(\xi_{k})\leq C_{u}$ is to ensure that the tasks assigned to the UAV cannot exceed its capacity during the time slot period. $y_{bt}(\xi_{k})\leq C_{b}x_{bt}$ gives the linear inequality relationship between continuous variable $y_{bt}(\xi_{k})$ and binary variable $x_{bt}$. If $x_{bt}=0$, and then $y_{bt}(\xi_{k})=0$. Otherwise, $y_{bt}(\xi_{k})$ must be less than or equal to the capacity of the selected BS $C_{b}$. A similar explanation is for $y_{st}(\xi_{k})\leq C_{s}x_{st}$. Constraint $(f)$ gives the equation of $\psi(\textbf{X},\textbf{Y},\xi_{k})$ appearing in the optimization objective, which represents the total system latency in $T$ time slots. It should be emphasized that in time slot $t$, UAV local processing, and additional BS server processing or cloud server remote processing are performed in parallel. Thus, we need to consider the maximum of the local processing latency and the offloading latency. Finally, constraint $(g)$ shows that the total energy consumption cannot be greater than the budget $E_{max}$. In general, optimization variables are $\textbf{X}$ and $\textbf{Y}$. The former one contains a series of binary variables, $\{x_{bt}, x_{st}, \forall  b,s,t\}$. The latter one is composed of $\{y_{ut}(\xi_{k}), y_{bt}(\xi_{k}), y_{st}(\xi_{k}), \forall  b,s,t,k\}$.

\section{Distributionally Robust Latency Optimization Algorithm Design}\label{S5}
In this section, we give a detailed distributionally robust latency optimization solution method to get the sub-optimal output and provide an analysis of the computational complexity.

\subsection{Algorithm Design}\label{S5-1}
For the formulated problem in (\ref{eq23}), we first interchange the minimization and maximization operations in the objective function, which does not change the nature of problem. Then, the optimization problem can be rewritten as
\begin{equation}\label{eq24}
\begin{split}
&\max\limits_{P\in D}\min\limits_{\textbf{X},\textbf{Y}}\mathbb{E}_{P}[\psi(\textbf{X},\textbf{Y},\xi_{k})]\\
& s.t. \\
& (\ref{eq23}a)\sim(\ref{eq23}g). \\
\end{split}
\end{equation}

To address the problem in (\ref{eq24}), we can temporarily treat the probability distribution $P$ as fixed, and then optimize the internal minimization problem. Since the formula of $\psi(\textbf{X},\textbf{Y},\xi_{k})$ is a bit complicated, we introduce additional variables $\{Q_{kt}, \forall  k,t\}$ to make some transformations. Then, the minimization problem is expressed as follows,
\begin{align}\label{eq25}
&\min\limits_{\textbf{X},\textbf{Y}}\mathbb{E}_{P}[\psi(\textbf{X},\textbf{Y},\xi_{k})]=\min\limits_{\textbf{X},\textbf{Y}}\sum\limits_{k=1}^{K}\sum\limits_{t=1}^{T}p_{k}\min Q_{kt}  \nonumber\\
=&\min\limits_{\textbf{X},\textbf{Y},Q_{kt}}\sum\limits_{k=1}^{K}\sum\limits_{t=1}^{T}p_{k}Q_{kt} \nonumber\\
& s.t. \nonumber\\
& (a)\ x_{bt} \in \{0,1\}, \ x_{st} \in \{0,1\}, \  \forall  b,s,t, \nonumber\\
& (b)\ \sum\limits_{b=1}^{N}x_{bt}+\sum\limits_{s=1}^{M}x_{st}=1, \  \forall  t, \nonumber\\
& (c)\ y_{ut}(\xi_{k})+\sum\limits_{b=1}^{N}y_{bt}(\xi_{k})+\sum\limits_{s=1}^{M}y_{st}(\xi_{k})=\xi_{k}, \ \forall  t,k, \nonumber\\
& (d)\ y_{ut}(\xi_{k}), y_{bt}(\xi_{k}), y_{st}(\xi_{k})\geq 0, \ \forall  b,s,t,k, \nonumber\\
& (e)\ y_{ut}(\xi_{k})\!\leq \!C_{u}, y_{bt}(\xi_{k})\!\leq \!C_{b}x_{bt}, y_{st}(\xi_{k})\!\leq \!C_{s}x_{st}, \forall  b,s,t,k, \nonumber\\
& (f)\  \frac{\delta y_{ut}(\xi_{k})}{f_{u}}\leq Q_{kt}, \  \forall  k,t, \nonumber\\
& (g)\  \sum\limits_{b=1}^{N}L_{ubt}+\sum\limits_{s=1}^{M}L_{ust}\leq Q_{kt},\  \forall  k,t, \nonumber \\
& (h)\ E^{fly}_{u}+\sum\limits_{b=1}^{N}\sum\limits_{t=1}^{T}E_{ubt} +\sum\limits_{s=1}^{M}\sum\limits_{t=1}^{T}E_{ust}\leq E_{max}, \ \forall  k.
\end{align}
This is a mixed integer programming problem, and we relax the discrete variables in matrix \textbf{X} to continuous variables ranging from 0 to 1. Then, the minimization problem can be transformed into
\begin{equation}\label{eq26}
\begin{split}
&\min\limits_{\textbf{X},\textbf{Y},Q_{kt}}\sum\limits_{k=1}^{K}\sum\limits_{t=1}^{T}p_{k}Q_{kt}\\
& s.t. \\
& (a)\ 0 \leq x_{bt}, x_{st} \leq 1, \  \forall  b,s,t, \\
& (\ref{eq25}b)\sim(\ref{eq25}h). \\
\end{split}
\end{equation}
However, we still have the maximization problem with the variable $P\in D$ in the outer layer. Thus, we transform the minimization problem into its dual problem shown as follows.
\begin{align}\label{eq27}
&\max\limits_{\lambda_{t},\tilde{\lambda}^{k}_{t}, \hat{\lambda}^{k}_{t}, \lambda_{k}, \mu^{k}_{t}, \mu^{k}_{bt}, \mu^{k}_{st},\nu^{k}_{t}}\sum\limits_{t=1}^{T}-\lambda_{t}+\sum\limits_{k=1}^{K}\lambda_{k}(E^{fly}_{u}-E_{max})\nonumber\\
&\ \ \ \ \ \ \ \  \ \ \ \ \ \ \ \  \ \ \ \ \ \ \  \  -\sum\limits_{t=1}^{T}\sum\limits_{k=1}^{K}\nu^{k}_{t}\xi_{k}
-\sum\limits_{t=1}^{T}\sum\limits_{k=1}^{K}\mu^{k}_{t}C_{u}\nonumber\\
& s.t. \nonumber\\
& (a)\ p_{k}-\tilde{\lambda}^{k}_{t}-\hat{\lambda}^{k}_{t}\geq 0, \  \forall  k,t, \nonumber\\
& (b)\ \lambda_{t}-\sum\limits_{k=1}^{K}\mu^{k}_{bt}C_{b}\geq 0, \  \forall  b,t, \nonumber\\
& (c)\ \lambda_{t}-\sum\limits_{k=1}^{K}\mu^{k}_{st}C_{s}\geq 0, \  \forall  s,t, \nonumber\\
& (d)\ \frac{\delta\tilde{\lambda}^{k}_{t}}{f_{u}}+\nu^{k}_{t}+\mu^{k}_{t} \geq 0,\  \forall  k,t,  \nonumber\\
& (e)\ \frac{\hat{\lambda}^{k}_{t}}{R_{ubt}}+\frac{\delta\hat{\lambda}^{k}_{t}}{f_{b}}+\frac{\varpi\hat{\lambda}^{k}_{t}}{R_{but}}+\frac{\lambda_{k}P_{ub}}{R_{ubt}}+
\mu^{k}_{bt}+\nu^{k}_{t}\geq 0, \ \forall  b,k,t, \nonumber\\
& (f)\ \frac{\hat{\lambda}^{k}_{t}}{R_{ust}}+\frac{\hat{\lambda}^{k}_{t}(1+\varpi)}{R_{sc}}+\frac{\delta\hat{\lambda}^{k}_{t}}{f_{c}}+\frac{\varpi\hat{\lambda}^{k}_{t}}{R_{sut}}
+\frac{\lambda_{k}P_{us}}{R_{ust}}+\frac{\lambda_{k}P_{s}}{R_{sc}} \nonumber\\
&\ \ \ \ \ +\mu^{k}_{st}+\nu^{k}_{t}\geq 0, \ \forall  s,k,t, \nonumber\\
& (g)\ \tilde{\lambda}^{k}_{t}, \hat{\lambda}^{k}_{t}, \lambda_{k}, \mu^{k}_{t}, \mu^{k}_{bt}, \mu^{k}_{st} \geq 0, \  \forall  b,s,k,t,
\end{align}
where $\lambda_{t}, \nu^{k}_{t}, \tilde{\lambda}^{k}_{t}, \hat{\lambda}^{k}_{t}, \lambda_{k}$ are dual variables corresponding to constraints (\ref{eq25}$b$), (\ref{eq25}$c$), (\ref{eq25}$f$), (\ref{eq25}$g$) and (\ref{eq25}$h$). Besides, $\mu^{k}_{t}, \mu^{k}_{bt}, \mu^{k}_{st}$ are dual variables corresponding to the three inequalities in constraint (\ref{eq25}$e$).

Combining the maximization dual problem and the outer maximization operation, we get the final reformulated form of the original optimization problem in (\ref{eq24}) as
\begin{align}\label{eq28}
&\max\limits_{p_{k},\lambda_{t},\tilde{\lambda}^{k}_{t}, \hat{\lambda}^{k}_{t}, \lambda_{k}, \mu^{k}_{t}, \mu^{k}_{bt}, \mu^{k}_{st},\nu^{k}_{t}}\sum\limits_{t=1}^{T}-\lambda_{t}+\sum\limits_{k=1}^{K}\lambda_{k}(E^{fly}_{u}-E_{max})\nonumber\\
&\ \ \ \ \ \ \ \ \ \ \  \ \ \ \ \ \ \ \  \ \ \ \ \ \ \  \ -\sum\limits_{t=1}^{T}\sum\limits_{k=1}^{K}\nu^{k}_{t}\xi_{k}
-\sum\limits_{t=1}^{T}\sum\limits_{k=1}^{K}\mu^{k}_{t}C_{u}\nonumber\\
& s.t. \nonumber\\
& (a)\ \sum\limits_{k=1}^{K}p_{k}=1, \nonumber\\
& (b)\ P\in D, \nonumber\\
& (\ref{eq27}a)\sim(\ref{eq27}g),
\end{align}
where constraint $(a)$ is the basic condition of probability distribution, and constraint $(b)$ indicates that the ambiguous distribution $P$ belongs to the confidence set $D$.

Different metrics correspond to different confidence sets. When $L_{1}$ norm and $L_{\infty}$ norm are chosen, $D$ in constraint $(\ref{eq28}b)$ is replaced by $D_{1}$ and $D_{\infty}$ shown in (\ref{eq10}) and (\ref{eq11}), respectively. Moreover, when studying the Kantorovich metric, although the confidence set $D_{kan}$ is also given, adding it to the constraint $(\ref{eq28}b)$ makes it extremely difficult to solve the problem. Thus, we need to transform the confidence set into an easy-to-solve formation. According to the definition of the Kantorovich metric as stated before, we can get the distance measure that matches our proposed problem, $\max\limits_{h_{k}}\sum\limits_{k=1}^{K}h_{k}p_{k}-\sum\limits_{k=1}^{K}h_{k}p^{0}_{k}$ with $|h_{x}-h_{y}|\leq \rho(\xi_{x},\xi_{y}), \xi_{x}\neq \xi_{y}$ in $\Omega$. For the part of $|h_{x}-h_{y}|$, we first construct a matrix $A_{L\times K}$ with $L$ rows and $K$ columns. For each row, we randomly select two columns, assigning one column to 1, and the other column to -1, so that $L=K*(K-1)$ combinations are obtained. Then, $|h_{x}-h_{y}|\leq \rho(\xi_{x},\xi_{y}), \xi_{x}\neq \xi_{y}$ can be rewritten as $\sum\limits_{k=1}^{K}a_{lk}h_{k}\leq b_{l}, \forall l=1,2,...,L$. Here, $b_{l}$ is the distance between $\xi_{x}$ and $\xi_{y}$ corresponding to two columns selected in the $l$-th row. What needs to be declared is that the physical meaning of $b_{l}$ is distance, not the absolute value of the difference between $\xi_{x}$ and $\xi_{y}$. Based on above analysis, about the confident set $D_{kan}$, we turn to consider the following problem,
\begin{equation}\label{eq29}
\begin{split}
&\max\limits_{h_{k}}\sum\limits_{k=1}^{K}h_{k}p_{k}-\sum\limits_{k=1}^{K}h_{k}p^{0}_{k}\\
& s.t. \\
& (a)\ \sum\limits_{k=1}^{K}a_{lk}h_{k}\leq b_{l}, \ \forall l=1,2,...,L.\\
\end{split}
\end{equation}

Then, we dualize problem (\ref{eq29}) to get the following formulation.
\begin{equation}\label{eq30}
\begin{split}
&\min\limits_{u_{l}}\sum\limits_{l=1}^{L}u_{l}b_{l} \\
& s.t. \\
& (a)\ \sum\limits_{l=1}^{L}u_{l}a_{lk}\geq p_{k}-p^{0}_{k}, \forall k=1,2,...,K, \\
& (b)\ u_{l}\geq 0, \forall l=1,2,...,L, \\
\end{split}
\end{equation}
where $u_{l}, \forall l=1,2,...,L$ is dual variable corresponding to the constraint (\ref{eq29}$a$). Finally, the following three constraints can be used to replace constraint (\ref{eq28}$b$) when the Kantorovich metric is applied.
\begin{equation}\label{eq31}
\sum\limits_{l=1}^{L}u_{l}b_{l}\leq \theta_{kan},
\end{equation}
\begin{equation}\label{eq32}
\sum\limits_{l=1}^{L}u_{l}a_{lk}\geq p_{k}-p^{0}_{k}, \forall k=1,2,...,K,
\end{equation}
\begin{equation}\label{eq33}
u_{l}\geq 0, \forall l=1,2,...,L.
\end{equation}

Now, we can solve problem (\ref{eq28}) using the Gurobi Optimizer and the worst case distribution $P$ is obtained. Then, going back to the original minimization problem in (\ref{eq26}) with $P$ substituted, the Gurobi Optimizer is also utilized to get $\textbf{X}$ and $\textbf{Y}$. Since the elements in $\textbf{X}$ are relaxed as continuous variables, we need to revert to integer variables of 0 or 1. Thus, the branch and bound algorithm is adopted. In this algorithm, we first select the branch variable, that is, which variable we want to recover from the relaxation value. In order to prune more nodes, our rule is to select the variable with the largest difference between the boundary and the relaxation value. Below we present the rule for selecting branch variables.
\begin{equation}\label{eq34}
x^{\ast}=\mathop{\arg\max}_{x_{bt},x_{st} \in \textbf{X}}\{\min\{1-x_{bt},x_{bt}\},\min\{1-x_{st},x_{st}\}\}.
\end{equation}
Then, we turn to branch node selection, that is, for the branch variable, whether to choose 0 or 1 for generating new optimization problem. We assume that $Lat0$ and $Lat1$ are lower bounds for node $x$ taking 0 and $x$ taking 1, respectively. The optimal result $Lat0$ can be obtained by adding $x\leq0$ as a constraint to problem (\ref{eq26}), and we add $x\geq1$ as a constraint of problem (\ref{eq26}) to output the optimal solution $Lat1$. Of course, if the problem is not feasible, we record the result as positive infinity. Then, we compare the value of $Lat0$ and $Lat1$. If $Lat0$ is less than $Lat1$, $x\leq0$ is finally selected to be added as the problem constraint. Otherwise, adding $x\geq1$ as a constraint is more effective. The above process of branch variable and branch node selections is repeated until all elements in $\textbf{X}$ are integers.

\begin{algorithm}[t]
\caption{Distributionally Robust Latency Optimization Algorithm}
\label{alg1}
\begin{algorithmic}[1]
\REQUIRE
  Sample space $\Omega$, reference distribution $P_{0}$ based on a series of historical data
\ENSURE
  $\textbf{X},\textbf{Y}$
\STATE Solve the optimization problem in (\ref{eq28}) by the Gurobi Optimizer to obtain $P=\{p_{1},p_{2},...,p_{K}\}$;
\STATE Solve the optimization problem in (\ref{eq26}) by the Gurobi Optimizer to get continuous $\textbf{X}$ and $\textbf{Y}$;
\REPEAT
\STATE  Select the branch variable $x^{\ast}$ according to the rule shown in (\ref{eq34});
\STATE  Add $x \leq 0$ as a constraint to problem (\ref{eq26}) and solve it to obtain the optimal result $Lat0$;
\STATE  Add $x \geq 1$ as a constraint to problem (\ref{eq26}) and solve it to obtain the optimal result $Lat1$;
\IF {$Lat0 < Lat1$}
  \STATE Update problem (\ref{eq26}) by adding a constraint $x \leq 0$;
\ELSE
  \STATE Update problem (\ref{eq26}) by adding a constraint $x \geq 1$;
\ENDIF
\UNTIL {all elements in $\textbf{X}$ are integers}.
\end{algorithmic}
\end{algorithm}

The details of the proposed distributionally robust latency optimization algorithm are summarized in Algorithm \ref{alg1}. First, we input the sample space $\Omega=\{\xi_{1},\xi_{2},...,\xi_{K}\}$. Then, given the size of historical data as $K^{\prime}$, a reference distribution $P_{0}$ is constructed. In step 1, no matter which of the three metrics is adopted, the maximization problem in (\ref{eq28}) is a convex problem, and we use a commercial optimization solver Gurobi to solve it. Then, the ambiguous distribution $P$ is obtained. In step 2, by substituting $P$ into the objective function of inner minimization problem (\ref{eq26}), it is a linear programming problem, and we also utilize Gurobi to output continuous decision variables $\textbf{X}$ and $\textbf{Y}$. Steps 3-12 describe the branch and bound algorithm, and finally continuous variables in $\textbf{X}$ are recovered to integer variables.

\subsection{Complexity Analysis}\label{S5-2}
In Algorithm \ref{alg1}, there are two optimization problems to be solved by the Gurobi Optimizer in steps 1 and 2. For problems (\ref{eq28}) and (\ref{eq26}), they contain $Q_{1}=T(5K+N+M+2NK+2MK)+K+1+N_{metric}$ and $Q_{2}=T(2N+2M+1+5K+2NK+2MK)+K$ linear constraints of size 1, respectively. $N_{metric}$ is the number of linear inequality and equality constraints introduced by the adopted metric. Then, the computational complexity of reaching $\varepsilon$-optimal solutions for these two problems is $\mathcal{O}(\sqrt{Q_{1}}\ln(1/\varepsilon)(\widehat{n}Q_{1}+\widehat{n}^{2}Q_{1}+\widehat{n}^{3})+\sqrt{Q_{2}}\ln(1/\varepsilon)(\widetilde{n}Q_{2}+\widetilde{n}^{2}Q_{2}+\widetilde{n}^{3}))$, where $\widehat{n}$ and $\widetilde{n}$ are decision variables with the order of $(2K+T+4KT+NKT+MKT)$ and $(NT+MT+KT)$, respectively \cite{Lu}. Moreover, since the UAV can only offload tasks to one server at a time slot, the branch and bound algorithm requires a maximum of $T(N+M-1)$ cycles to solve problem (\ref{eq26}), and each cycle involves a performance comparison of a branch variable with two branch nodes selected. From the first cycle to the $T(N+M-1)$-th cycle, the number of constraints for problem (\ref{eq26}) also increases from $Q_{2}+1$ to $Q_{2}+T(N+M-1)$ with a step size of 1. In the end, for the branch and bound algorithm, a total of $2T(N+M-1)$ problems need to be solved by the Gurobi Optimizer, and the complexity is $\mathcal{O}\Bigg(\sum\limits_{i=1}^{T(N+M-1)}2(\sqrt{(Q_{2}+i)}\ln(1/\varepsilon)(\widetilde{n}(Q_{2}+i)+\widetilde{n}^{2}(Q_{2}+i)+\widetilde{n}^{3}))\Bigg)$. Thus, the complexity of the distributionally robust latency optimization algorithm is $\mathcal{O}\Bigg(\sqrt{Q_{1}}\ln(1/\varepsilon)(\widehat{n}Q_{1}+\widehat{n}^{2}Q_{1}+\widehat{n}^{3})+\sqrt{Q_{2}}\ln(1/\varepsilon)(\widetilde{n}Q_{2}+\widetilde{n}^{2}Q_{2}+\widetilde{n}^{3})+
\sum\limits_{i=1}^{T(N+M-1)}2(\sqrt{(Q_{2}+i)}\ln(1/\varepsilon)(\widetilde{n}(Q_{2}+i)+\widetilde{n}^{2}(Q_{2}+i)+\widetilde{n}^{3}))\Bigg)$.

\section{Performance Evaluation}\label{S6}
In this section, compared with the traditional deterministic optimization and other representative schemes, we evaluate the robustness and effectiveness of the proposed algorithm. Then, different metrics used to construct the confidence set are analyzed under various system parameters.
\begin{table}[t]
\begin{center}
\caption{SIMULATION PARAMETERS}
\begin{tabular}{ccc}
\hline
Parameter & Symbol  & Value \\
\hline
Noise spectral density &  $N_{0}$  & -174dBm/Hz \\
Rician factor & $\eta$ & 7\\
UAV capacity & $C_{u}$ & 3Mbps \\
BS capacity&  $C_{b}$ & 25Mbps \\
Cloud capacity  &  $C_{s}$   &  150Mbps \\
Transmit power for UAV$\rightarrow$ BS &  $P_{ub}$  & 1.6W \\
Transmit power for BS$\rightarrow$ UAV &  $P_{bu}$  & 1.6W \\
Transmit power for UAV$\rightarrow$ satellite &  $P_{us}$  & 5W \\
Transmit power for satellite$\rightarrow$ UAV &  $P_{su}$  & 5W \\
Transmit power for satellite &  $P_{s}$  & 5W \\
CPU processing speed for UAV &   $f_{u}$ &    $3*10^8$cycles/s \\
CPU processing speed for BS &   $f_{b}$ &    $5*10^9$cycles/s \\
CPU processing speed for cloud server &   $f_{c}$ &    $10*10^9$cycles/s \\
Transmission rate for satellite to cloud  &   $R_{sc}$   & 20Mbps \\
Antenna gain &   $G_{0}$  &    43.3dBi \\
\hline
\end{tabular}
\label{table1}
\end{center}
\end{table}
\subsection{Simulation Setup}\label{S6-1}
In the investigated SAGIN system, we take the UAV's initial location at $H=100$m above the center of the remote area. It flies along a circular trajectory centered at $(1000, 0, 100)$m and with a radius of $r=1000$m. The flying speed is $1$km/min, $c_{1}=9.26*10^{-4}$ and $c_{2}=2250$. Multiple nearby BSs are deployed in a rectangular area with four points as vertices, $(500, -1000, 100)$m, $(2000, -1000, 100)$m, $(2000, 1000, 100)$m and $(500, 1000, 100)$m. In each time slot, we assume that the UAV always has a relatively close BS that can be connected. Moreover, several LEO satellites are obtained by the Systems Tool Kit. With reference to the Iridium constellation system, the constellation we construct has 2 orbital planes and each with 4 satellites. The altitudes vary from $780$km to $800$km, and the position of the satellite changes dynamically in different time slots. In addition, for the large-scale fading of the UAV and BS channel, we use the UMi path loss model \cite{UMi}, and the bandwidth of C-band is 20MHz. While the free space path loss model is adopted to express the large-scale channel fading between the UAV and satellite, and $B_{us}=400$MHz. For different applications, the amount of computing data and the returned data size are also different. In this paper, $\delta=25$cycles/bit and $\varpi=1/1000$. Finally, we use the data set from a real-world network of the SNDLib library \cite{SNDlib}. The data set provides 161 nodes, corresponding coordinates and demand values of multiple data streams. It spans 8 days and the granularity is $1$min. $\tau=1$min. Combined with the system model, we only study the task volume in the first 5 hours of each day. The number of basic events in the sample space is set as $K=9$, and we use $10\%$ of the sum of demand values per minute as the task volume collected by the UAV in current time slot. The other system parameters are shown in Table \ref{table1}.

To prove the superiority of the proposed algorithm, the name of which is replaced by the used metric and we compare it with the following algorithms.
\begin{itemize}
\item \textbf{Greedy Algorithm}: for this algorithm, we simply write by Greedy in the following. It is also assumed that the probability distribution of arriving task volume is uncertain, and the UAV selects a BS or LEO satellite for connection in each time slot. The difference is that we allow the UAV to offload all collected tasks to the edge server or the cloud server with considerable capacity for processing.
\item \textbf{Deterministic Optimization}: this algorithm, labeled as Deterministic, is the traditional deterministic optimization algorithm. Although the amount of tasks arriving is uncertain, and we have no information about its probability distribution, the UAV estimates it to a certain value for designing offloading strategy. Here, the average value of basic events in sample space is used as the determined estimate.
\item \textbf{Greedy Deterministic Optimization}: different from the Deterministic Optimization, this scheme is to offload all determined estimated tasks to the server for processing, and the UAV itself does not perform local computing. Later, we will use the name of Greedy deterministic instead.
\end{itemize}

\begin{figure}[!t]
\begin{center}
\includegraphics*[width=0.7\columnwidth,height=1.9in]{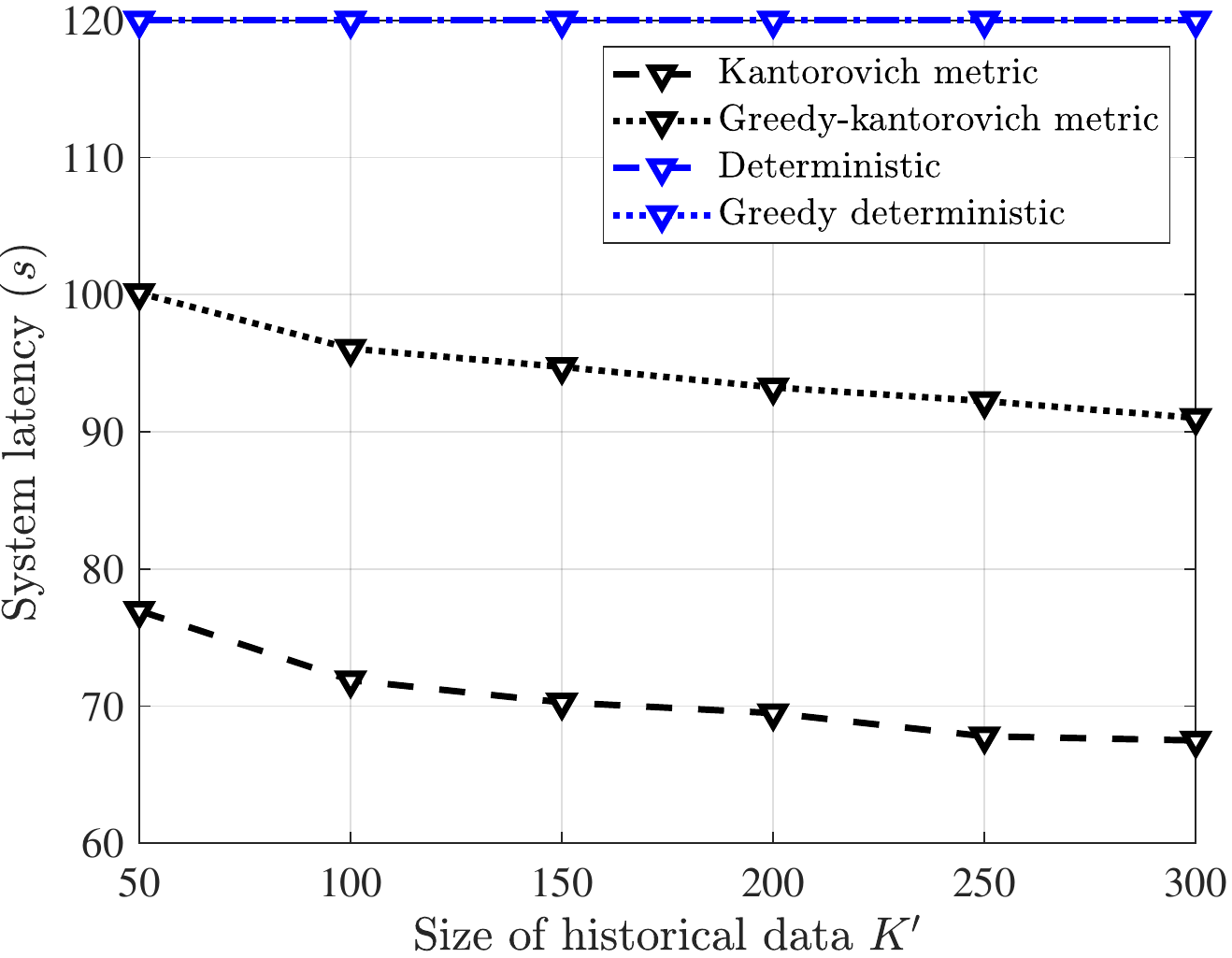}
\end{center}
\caption{System latency v.s. size of historical data for four schemes.} \label{fig2}
\end{figure}

\begin{figure*}[!t]
\begin{minipage}[t]{0.5\linewidth}
\centering
\includegraphics[width=0.7\columnwidth,height=1.9in]{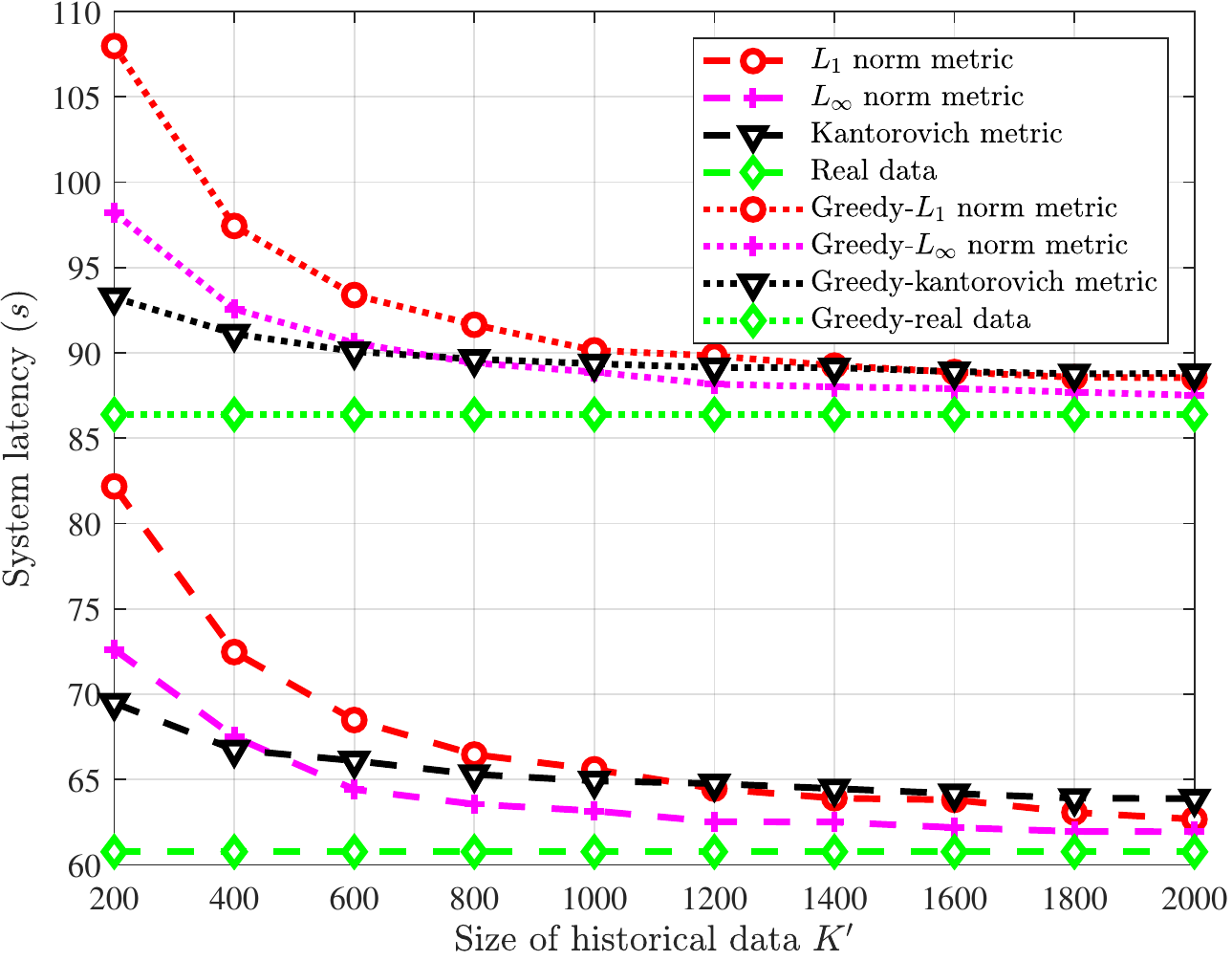}
\centerline{\small (a)}
\end{minipage}%
\begin{minipage}[t]{0.5\linewidth}
\centering
\includegraphics[width=0.7\columnwidth,height=1.9in]{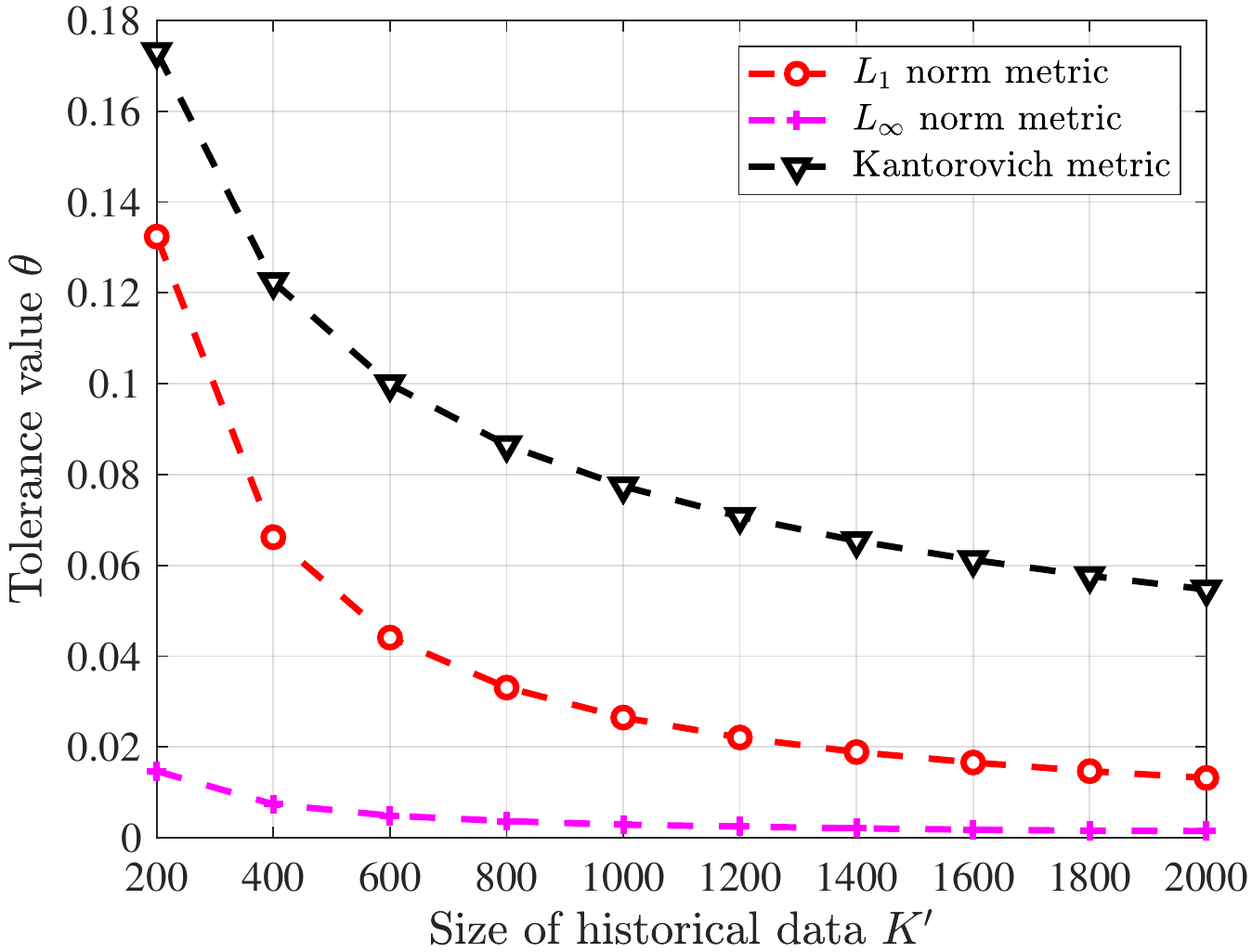}
\centerline{\small (b)}
\end{minipage}
\begin{minipage}[t]{0.5\linewidth}
\centering
\includegraphics[width=0.7\columnwidth,height=1.9in]{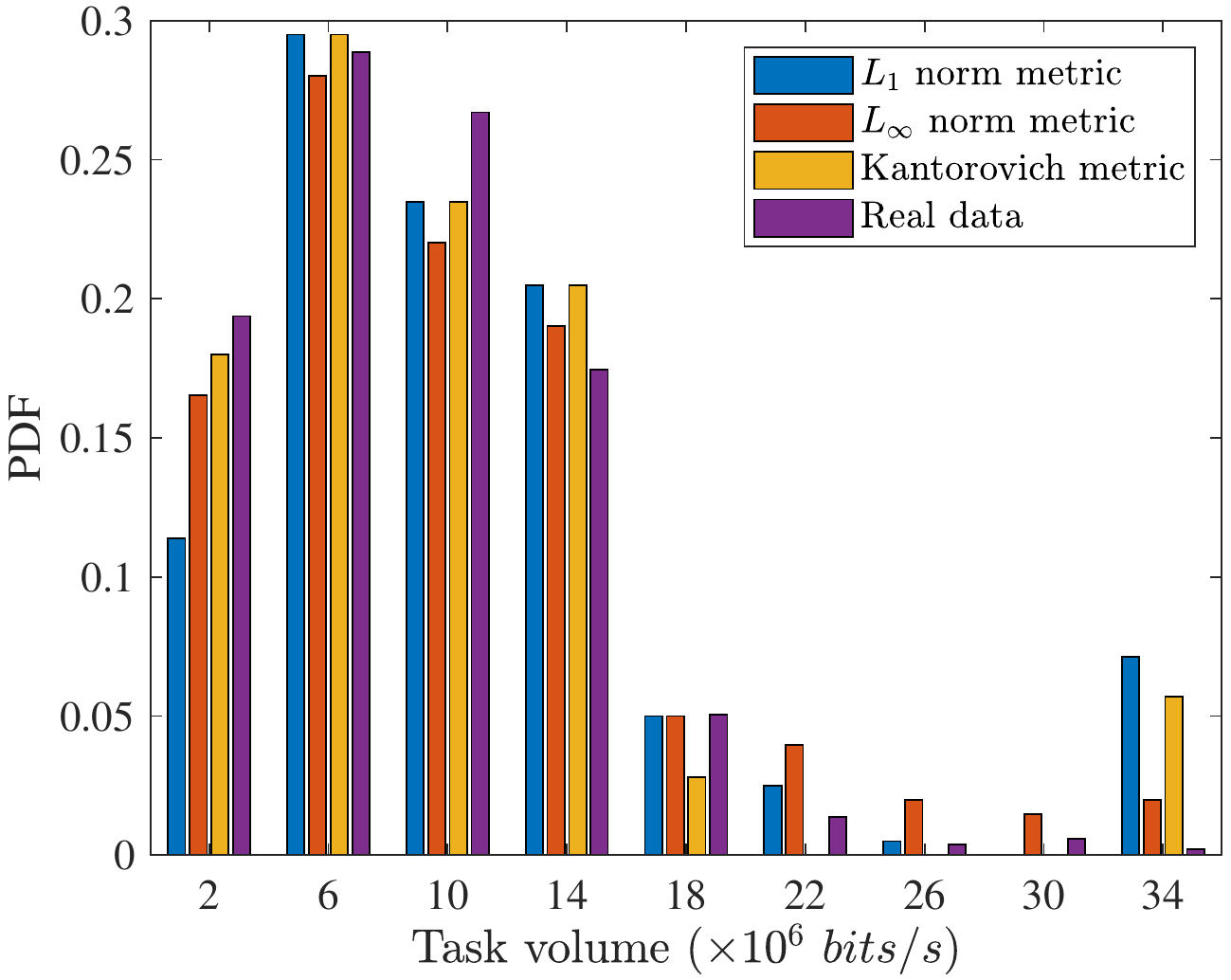}
\centerline{\small (c)}
\end{minipage}
\begin{minipage}[t]{0.5\linewidth}
\centering
\includegraphics[width=0.7\columnwidth,height=1.9in]{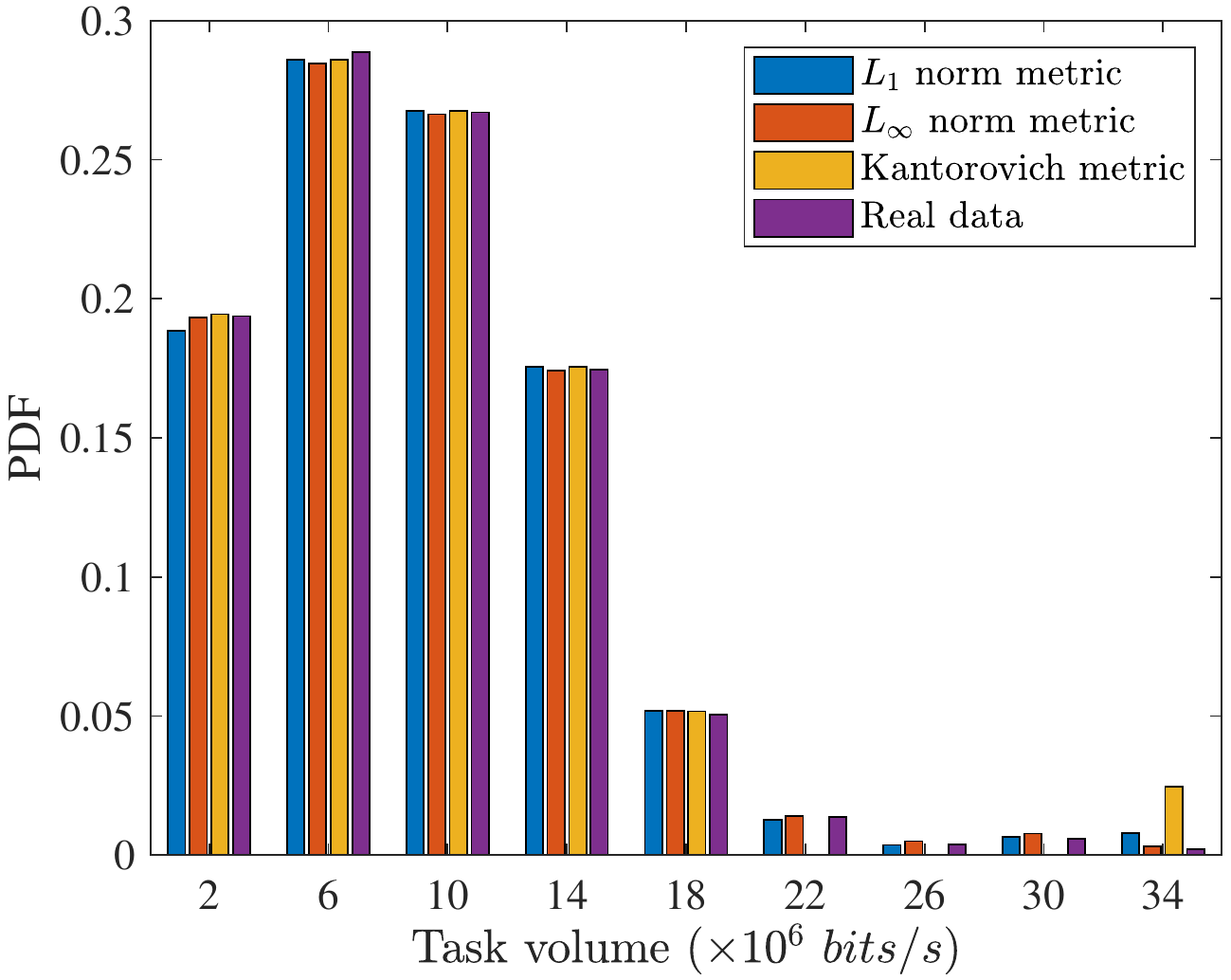}
\centerline{\small (d)}
\end{minipage}
\caption{(a) System latency v.s. size of historical data $K^{\prime}$ for different metrics; (b) Tolerance value $\theta$ v.s. size of historical data $K^{\prime}$; (c) PDF v.s. task volume with the size of historical data $K^{\prime}=200$; (d) PDF v.s. task volume with the size of historical data $K^{\prime}=2200$.}
\vspace*{-3mm}
\label{fig3}
\end{figure*}

\subsection{Performance Analysis}\label{S6-2}
In Fig. \ref{fig2}, we set the number of BSs and LEO satellites to be 5 and 3. The confident level is 0.95 and $T=2$. If there is no special statement later, the values of these parameters will not change. Then, by using the Kantorovich metric to construct the confident set, we give a detailed comparison about the system latency of four schemes with the increase of historical data size from 50 to 300. We can see that the proposed algorithm achieves the best system performance, followed by the greedy algorithm, and the other two schemes are the worst. For two schemes, Deterministic and Greedy deterministic, because the average value is taken as a certain default value, when there is a serious deviation between the arrived task volume and the estimated value, task drop will occur and retransmission is required. For instance, if the UAV selects one BS to access according to the estimated value, and the task volume far exceeds the BS capacity, which will result in greatly increased task drop rate. In this paper, we record the retransmission latency as the length of the time slot, 60s. In general, for the optimization under uncertainty, the proposed algorithm indeed avoids some shortcomings of traditional deterministic optimization and can adjust decisions flexibly according to the observed portion of data. Thus, the system robustness is greatly improved. According the results, the system latency of Algorithm \ref{alg1} is reduced by $44.6\%$ and $27.8\%$ in comparison with the deterministic optimization and greedy algorithm when $K^{\prime}=300$, respectively.

In Fig. \ref{fig3}(a), for the proposed algorithm and the greedy algorithm, we show the impact of three metrics on the system performance. The Real data scheme corresponds to the system latency under the real distribution generated by the data in the adopted data set. As the size of historical data increases and more information is available, no matter which metric is applied, the system latency will gradually decrease and become closer to the system latency under the real distribution. Besides, for two metrics of the same type, $L_{1}$ norm and $L_{\infty}$ norm, the latter one is more effective. For the Kantorovich metric belonging to the $\zeta$-structure metrics family, it notably outperforms others when the amount of historical data obtained is small. However, when the observed data size is large, its performance is not as good as that of $L_{\infty}$ norm, and the Kantorovich metric is even slightly inferior to $L_{1}$ norm. Generally, the performance under $L_{1}$ norm and $L_{\infty}$ norm metrics is more affected by the size of historical data. In Fig. \ref{fig3}(b), under these three metrics, curves of the tolerance value $\theta$ with the historical data size are presented. Also, we compare metrics of the same type first. It indicates that $L_{\infty}$ norm metric has the smaller $\theta$ than $L_{1}$ norm, and it is less conservative. It can be explained by the fact that the confidence set shrinks as $\theta$ becomes smaller, and then the conservativeness will decrease. These results are also consistent with the effectiveness analysis in Fig. \ref{fig3}(a). Then, it can be readily found that the Kantorovich metric has the highest conservativeness and the system reliability is better. Meanwhile, it is very beneficial to maintain a better system performance especially when there is less data information. Fig. \ref{fig3}(a) and Fig. \ref{fig3}(b) provide a reference for employing metrics. Fig. \ref{fig3}(c) and Fig. \ref{fig3}(d) intuitively show the ambiguous distribution estimated to replace the true distribution for different metrics when the historical data sizes are 200 and 2200 respectively. We can conclude that when the amount of data is large, the estimated ambiguous distribution is very close to the real distribution, especially for $L_{1}$ norm and $L_{\infty}$ norm. While the data size is small, the estimated results still have some deviations.

\begin{figure}[!t]
\begin{center}
\includegraphics*[width=0.7\columnwidth,height=1.9in]{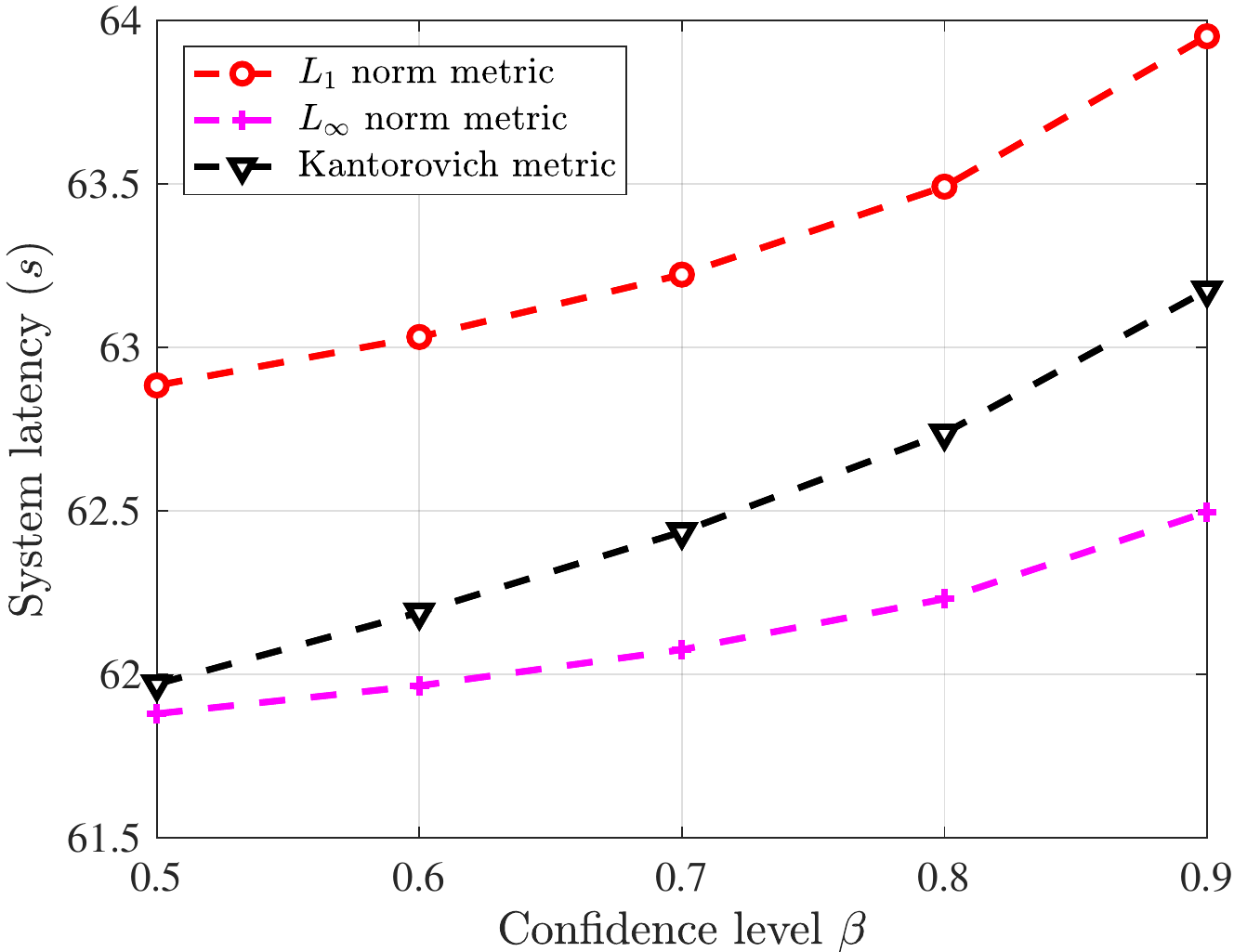}
\end{center}
\caption{System latency v.s. confidence level $\beta$.} \label{fig4}
\end{figure}
In Fig. \ref{fig4}, we use 1000 data to form a reference distribution and plot the system latency with the confidence level varied from 0.5 to 0.9 with step size of 0.1. For different metrics that match the proposed algorithm, it is worth mentioning that the larger the $\beta$, the larger the $\theta$ as a function of $\beta$, the higher the system reliability, and the higher the conservativeness, which leads to a decrease in effectiveness and an increase in system latency.

\begin{figure}[!t]
\begin{center}
\includegraphics*[width=0.7\columnwidth,height=1.9in]{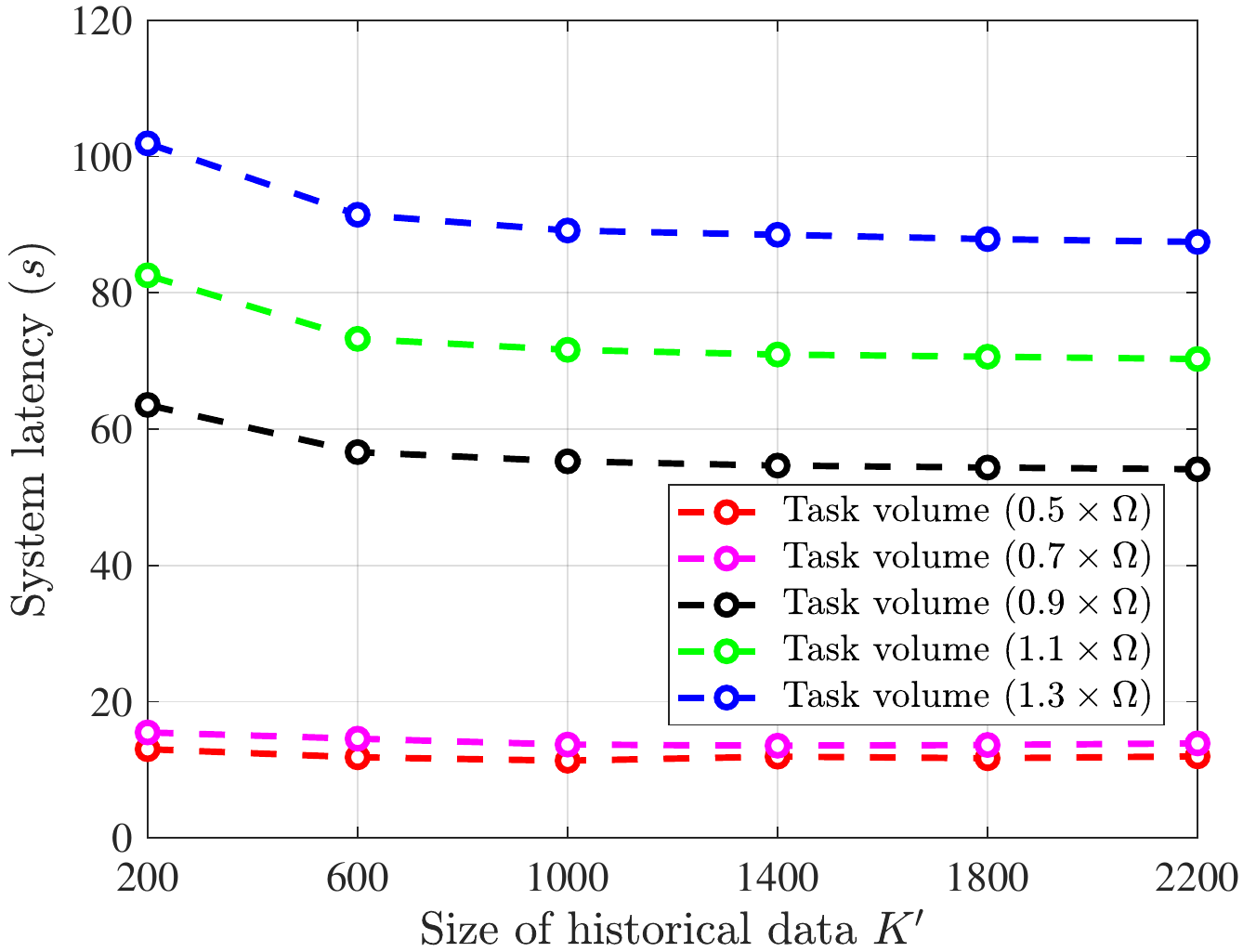}
\end{center}
\caption{System latency v.s. size of historical data under different task volumes.} \label{fig5}
\end{figure}

In Fig. \ref{fig5}, we take different sample spaces as $\Omega$ multiplied by $0.5$, $0.7$, $0.9$, $1.1$, and $1.3$. Then, Fig. \ref{fig5} depicts the system latency changes with the size of historical data under different sample spaces of arrival tasks. As the amount of tasks increases and exceeds the BS capacity boundary, the UAV will choose LEO satellites and the latency shows a sharp increase. Then, one can also remark that when the task volume is small, such as red and magenta curves, the capacity of the BS is sufficient, and then the UAV selects a nearby BS to offload tasks. In general, ground BSs have made a great contribution to reduce the system latency, while LEO satellites have a significant advantage in capacity.

\begin{figure}[!t]
\begin{center}
\includegraphics*[width=0.7\columnwidth,height=1.9in]{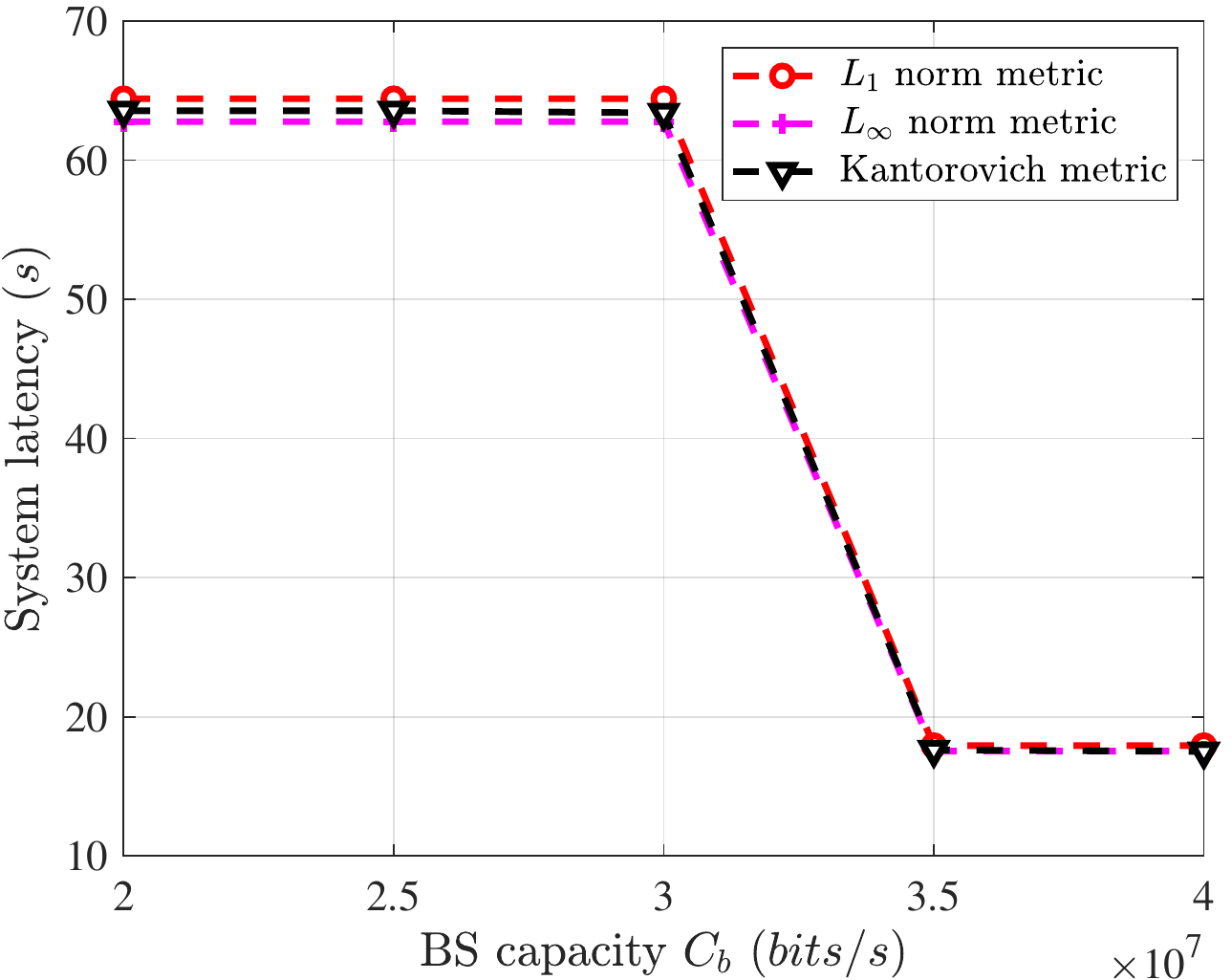}
\end{center}
\caption{System latency v.s. BS capacity $C_{b}$.} \label{fig6}
\end{figure}

Fig. \ref{fig6} illustrates the system latency v.s. BS capacity. It is observed that for the same task volume sample space, when the BS capacity reaches a certain critical value, the UAV will switch to the BS for communication and computing offloading. As expected, latency overhead is rapidly reduced and the system performance is improved.
\section{Conclusion}\label{S7}

In this paper, we have used the SAGIN architecture to serve a large number of IoT devices in the remote area. The UAV collects dynamically updated tasks above the interested area with preset trajectory, and then offloads a part of them to a BS or the satellite to utilize high-capacity server for processing. Since the probability distribution of the task volume is unknown, we have formulated a problem of minimizing the expectation of system latency under the worst-case distribution, and adopted the distributionally robust latency optimization algorithm to yield the sub-optimal solution. Numerical results have validated the superiority and robustness of the proposed algorithm in comparison with the greedy algorithm and the deterministic optimization algorithm. Besides, when the size of historical data is very small, the most conservative Kantorovich metric has the best performance, and when the size of historical data is large, the least conservative $L_{\infty}$ norm metric has the best performance. In the future work, we will study the space-air-ground integrated high-speed railway network scenario.

\bibliographystyle{IEEEtran}

\end{document}